\documentclass[preprint,amsmath,amssymb,superscriptaddress,longbibliography,aps,floatfix,footinbib]{revtex4-2}
\usepackage{graphicx}
\usepackage{tabularx}
\usepackage[usenames,dvipsnames]{color}
\usepackage{soul}
\usepackage{bm}
\usepackage{amsmath}
\usepackage{lineno}
\usepackage{gensymb}

\usepackage{times}
\usepackage{amsfonts}
\usepackage{mathrsfs}
\usepackage{graphicx}
\usepackage{dcolumn}
\usepackage{bm}
\usepackage{color}

\usepackage[colorlinks,bookmarks=false,citecolor=blue,linkcolor=red,urlcolor=blue]{hyperref}
\usepackage{multirow}
\usepackage{physics}

\begin{document}

\title{Bilayer orthogonal ferromagnetism in CrTe$_2$-based van der Waals system}

\author{Chiara Bigi**}
\email{chiara.bigi@synchrotron-soleil.fr}
\affiliation{Synchrotron SOLEIL, F-91190 Saint-Aubin, France}

\author{Cyriack Jego**}
\affiliation{Univ. Grenoble Alpes, CEA, CNRS, Grenoble INP, IRIG-SPINTEC, 38000 Grenoble, France}

\author{Vincent Polewczyk**}
\affiliation{Univ. Grenoble Alpes, CEA, CNRS, Grenoble INP, IRIG-SPINTEC, 38000 Grenoble, France}
\affiliation{Universit\'e Paris-Saclay, UVSQ, CNRS, GEMaC, 78000, Versailles, France}

\author{Alessandro De Vita}
\affiliation{Fritz Haber Institut der Max Planck Gesellshaft, Faradayweg 4--6, 14195 Berlin, Germany\looseness=-1}

\author{Thomas Jaouen}
\affiliation{Univ Rennes, IPR Institut de Physique de Rennes, UMR 6251, F-35000 Rennes, France}

\author{Hulerich C. Tchouekem}
\affiliation{Univ Rennes, IPR Institut de Physique de Rennes, UMR 6251, F-35000 Rennes, France}

\author{Fran\c cois Bertran}
\affiliation{Synchrotron SOLEIL, F-91190 Saint-Aubin, France}

\author{Patrick Le F\`evre}
\affiliation{Univ Rennes, IPR Institut de Physique de Rennes, UMR 6251, F-35000 Rennes, France}

\author{Pascal Turban}
\affiliation{Univ Rennes, IPR Institut de Physique de Rennes, UMR 6251, F-35000 Rennes, France}

\author{Jean-Fran\c cois Jacquot}
\affiliation{Univ. Grenoble Alpes, CEA, CNRS, IRIG-SYMMES, 38000 Grenoble, France}

\author{Jill A. Miwa}
\affiliation{Department of Physics and Astronomy, Interdisciplinary Nanoscience Center, Aarhus University, 8000 Aarhus C, Denmark}

\author{Oliver J. Clark}
\affiliation{School of Physics and Astronomy, Monash University, Clayton, Victoria 3800, Australia}

\author{Anupam Jana}
\affiliation{CNR-IOM Istituto Officina dei Materiali, I-34139 Trieste, Italy}

\author{Sandeep Kumar Chaluvadi}
\affiliation{CNR-IOM Istituto Officina dei Materiali, I-34139 Trieste, Italy}

\author{Pasquale Orgiani}
\affiliation{CNR-IOM Istituto Officina dei Materiali, I-34139 Trieste, Italy}

\author{Mario Cuoco}
\affiliation{CNR-SPIN, c/o Universit\'a di Salerno, IT-84084 Fisciano (SA), Italy}

\author{Mats Leandersson}
\affiliation{MAX IV Laboratory, Lund University, Lund, Sweden}

\author{Thiagarajan Balasubramanian}
\affiliation{MAX IV Laboratory, Lund University, Lund, Sweden}

\author{Thomas Olsen}
\email{tolsen@fysik.dtu.dk}
\affiliation{CAMD, Computational Atomic-Scale Materials Design, Department of Physics, Technical University of Denmark, 2800, Kongens Lyngby, Denmark}

\author{Younghun Hwang}
\email{younghh@uc.ac.kr}
\affiliation{Electricity and Electronics and Semiconductor Applications, Ulsan College, Ulsan 44610, Republic of Korea}

\author{Matthieu Jamet}
\email{matthieu.jamet@cea.fr}
\affiliation{Univ. Grenoble Alpes, CEA, CNRS, Grenoble INP, IRIG-SPINTEC, 38000 Grenoble, France}

\author{Federico Mazzola}
\email{federico.mazzola@spin.cnr.it}
\email{federico.mazzola.15@gmail.com}
\affiliation{Department of Molecular Sciences and Nanosystems, Ca Foscari University of Venice, Venice IT-30172, Italy}
\affiliation{CNR-SPIN UOS Napoli, Complesso Universitario di Monte Sant’Angelo, Via Cinthia 80126, Napoli, Italy}

\begin{abstract}

\textbf{Systems with pronounced spin anisotropy play a pivotal role in advancing magnetization switching and spin-wave generation mechanisms, which are fundamental for spintronic technologies. Quasi-van der Waals ferromagnets, particularly Cr$_{1+\delta}$Te$_2$ compounds, represent seminal materials in this field, renowned for their delicate balance between frustrated layered geometries and magnetism. Despite extensive investigation, the precise nature of their magnetic ground state, typically described as a canted ferromagnet, remains contested, as does the mechanism governing spin reorientation under external magnetic fields and varying temperatures. In this work, we leverage a multimodal approach, integrating complementary techniques, to reveal that Cr$_{1+\delta}$Te$_2$ ($\delta = 0.25-0.50$) hosts a previously overlooked magnetic phase, which we term orthogonal-ferromagnetism. This single phase consists of alternating atomically sharp single layers of in-plane and out-of-plane ferromagnetic blocks, coupled via exchange interactions and as such, it differs significantly from crossed magnetism, which can be achieved exclusively by stacking multiple heterostructural elements together. Contrary to earlier reports suggesting a gradual spin reorientation in CrTe$_2$-based systems, we present definitive evidence of abrupt spin-flop-like transitions. This discovery, likely due to the improved crystallinity and lower defect density in our samples, repositions Cr$_{1+\delta}$Te$_2$ compounds as promising candidates for spintronic and orbitronic applications, opening new pathways for device engineering.}
\end{abstract}
\maketitle

** First Authors Shared - These authors contributed equally\\

Cr$_{1+\delta}$Te$_2$ transition metal dichalcogenides (TMDs) have rapidly emerged as a key platform for exploring magnetism in low-dimensional systems, offering unprecedented opportunities for discovering novel quantum phases \cite{doi:10.1021/acs.nanolett.4c01005, Fujisawa2023, Fujisawa2020, PhysRevMaterials.7.054005, 10.1063/5.0200063, Freitas_2015, doi:10.1021/acsami.0c07017, Zhang2021, Dijkstra_1989, doi:10.1021/acsanm.9b01179, ZHOU2022162223, 10.1063/5.0068018, 10.1063/5.0070079, PRAMANIK201772, doi:10.1021/acs.nanolett.1c02940, doi:10.1021/acs.nanolett.9b02191}. The magnetic properties of Cr-based dichalcogenides, particularly Cr$_{1.25}$Te$_2$, represent a complex and unresolved area of study. This phase, stabilized by 25$\%$ excess Cr, exhibits magnetic behavior that remains poorly understood. Diverging interpretations in the literature highlight the subtle nature of its magnetism. Tang et al. characterized Cr$_{1.25}$Te$_2$ as a ferromagnet with a pronounced out-of-plane spin polarization, indicating well-defined magnetic order \cite{Tang_2022}. In contrast, Liu et al. identified the same material as exhibiting a canted-ferromagnetic structure \cite{PhysRevB.100.245114}.\\

Discrepancies also emerge when considering the dynamic behavior of spins under external magnetic fields and varying temperatures. While a continuous reordering of spins has been observed, a downturn in magnetization raises questions about the underlying thermodynamic processes \cite{PhysRevB.100.245114}. Such conflicting behaviors suggest that Cr$_{1.25}$Te$_2$ exhibits a heightened sensitivity to perturbations, such as Cr vacancies. Further complications arise from the several magnetic transition temperatures reported, making efforts to define the properties of the idealized trigonal phase. This variability suggests that the observed magnetic behavior may be modulated by subtle structural factors, influencing the system's response to external stimuli. As a result, pure trigonal Cr$_{1.25}$Te$_2$ emerges as an exceptionally intriguing system—not only for its complex magnetic ordering but also for the underlying physical mechanisms governing the thermodynamic behavior of the spins. Unraveling these complexities would mark a significant advance in understanding magnetism in TMDs, potentially uncovering new magnetic phenomena in related systems. Thus, the focus of this work is to address these unresolved challenges.\\

By synthesizing high-purity single crystals of Cr$_{1.25}$Te$_2$ in the trigonal phase, we employ a comprehensive multimodal approach, combining superconducting quantum interference device (SQUID) magnetometry, spin- and angle-resolved photoelectron spectroscopy (Spin-ARPES), and density functional theory (DFT) calculations, to uncover a novel magnetic ground state, which we term orthogonal ferromagnetism. This emergent phase is defined by the coexistence of alternating in-plane and out-of-plane ferromagnetic moments, coupled through a significant antiferromagnetic exchange interaction. The new magnetic order observed differs in nature from crossed magnetism, exclusively achievable by stacking multiple heterostructural elements together or grown as thin-films \cite{Vincent1, Vincent2,Vincent3,Vincent4,Vincent5}. Indeed, orthogonal magnetism manifests as a single phase with alternating ferromagnetic stacking at the atomic scale and interfacial problems of typical crossed magnets can be completely surpassed. The system undergoes a pronounced, abrupt reordering, strikingly reminiscent of spin-flop transitions, marking a clear departure from the thermodynamic behaviour previously observed. Remarkably, this orthogonal ferromagnetic state demonstrates exceptional robustness, persisting across Cr doping levels up to at least 50$\%$. These findings resolve longstanding discrepancies in the literature concerning the magnetic behavior of Cr$_{1+\delta}$Te$_2$, recontextualizing the material as a promising platform for future magnetic switching technologies.\\

\begin{figure}
\centering
    \includegraphics[width=.8\textwidth]{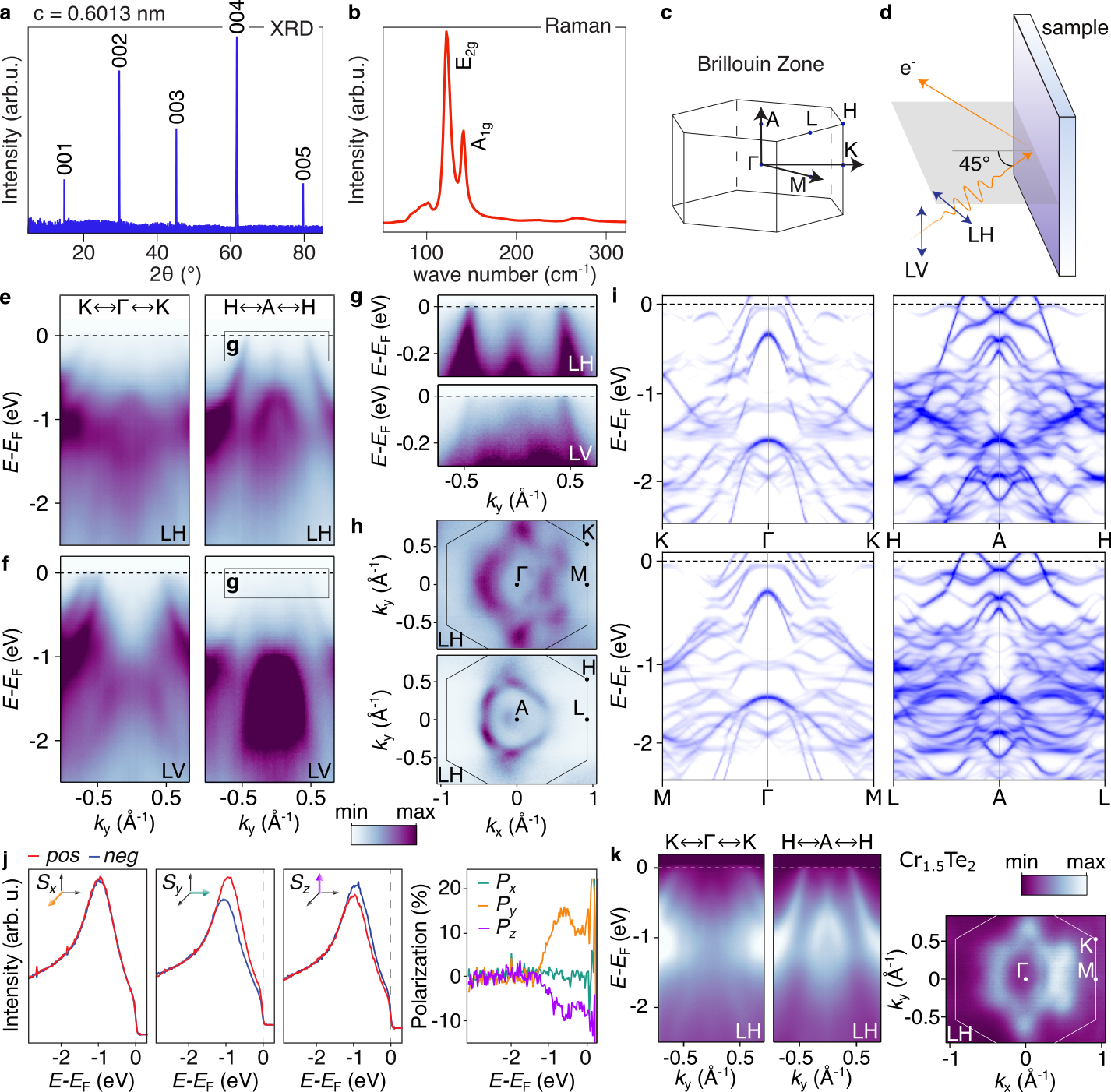}
    \caption{{\bf Crystalline and electronic structure of Cr$_{1+\delta}$Te$_{2}$} {\bf a.} The \textit{c}-axis lattice parameter obtained by x-ray diffraction (XRD) on high-quality single crystals is consistent with $\delta=0.25$ \cite{Purbawati_2024}. {\bf b.} Raman spectroscopy measurements reporting the $E_{2g}$ and $A_{1g}$ phonon manifolds with narrow lines, further corroborates the 1.25:2 stoichiometry \cite{FU2021122222}. {\bf c.} Brillouin zone {\bf d.} Cartoon of the experimental geometry and light polarization vectors used for ARPES measurements {\bf e.} Cr$_{1.25}$Te$_{2}$ Band dispersions  acquired along the ${\Gamma}-{K}$ (h$\nu=60$ eV) and ${A}-{H}$ (h$\nu=80$ eV) high symmetry directions with linear horizontal light polarization. {\bf f.} Same set probed with linear vertical light polarization to uncover the orbital mixing at the Fermi level. {\bf g.} The electron pocket at the Fermi level is well captured by our ab-initio calculations and allowed us to estimate a $V_0=8$ eV. {\bf h.} Fermi surface maps probed with (\emph{top})$60$ eV and (\emph{bottom}) $80$ eV photon energies, corresponding to the bulk Brillouin zone centre ($\Gamma-K-M$) and the Brillouin zone boundary ($A-H-L$), respectively. {\bf i.} Band structure calculations along $\Gamma-K$, $\Gamma-M$, $A-H$ and $A-L$ high symmetry directions. {\bf j.} Vectorial micro-spin spectroscopy performed at the $\Gamma$ point (h$\nu=100$ eV) the resulting polarization vector is consistent with the magnetic ground state of our model. {\bf k.} Electronic structure of the Cr$_{1.5}$Te$_{2}$ sister compound which displays similar magnetic phenomenology.}
    \label{fig1}
\end{figure}

The single crystals of Cr$_{1.25}$Te$_2$ were synthesized via the vertical temperature gradient Bridgman method, employing ultra-high purity Cr (99.99 $\%$) and Te (99.99999 $\%$) as starting materials (Cr$_{1.50}$Te$_2$ was also grown using the same methodology; the phase and composition were checked via XRD and X-ray core level spectroscopy, respectively). The stoichiometry of the grown crystals was verified through energy dispersive X-ray analysis, confirming the precise Cr content. Detailed procedural steps are described in the Methods section. Structural characterization demonstrated excellent agreement with other works \cite{Purbawati_2024}. Notably, XRD analysis revealed a \textit{c}-axis lattice parameter of Cr$_{1.25}$Te$_2$ as large as 0.6013 nm—slightly contracted relative to pure CrTe$_2$ \cite{Freitas_2015}—suggesting a clear structural impact from Cr intercalation and enhanced interlayer coupling (Fig.\ref{fig1} a). The Raman spectra on this system exhibited pronounced $E{2g}$ and $A{1g}$ phonon modes, affirming the high crystallinity and quality of the synthesized crystals (Fig.\ref{fig1} b).\\

To resolve the low-temperature magnetic ground state of such systems, we integrated structural insights from X-ray diffraction (XRD) and Raman spectroscopy with first-principles density functional theory (DFT) calculations (see Methods). DFT parameters were meticulously optimized to reproduce the experimentally observed electronic structure at 15–20 K, as revealed by angle-resolved photoemission spectroscopy (ARPES, Fig.\ref{fig1} e-i). This stringent alignment between theory and experiment enabled a definitive determination of the magnetic moment configurations, which are vital for accurately capturing the observed phenomena. Non-magnetic calculations, detailed in the Supplementary Information (see supplementary figure 1), were inconsistent with experimental results, reinforcing the necessity of incorporating magnetic order. Furthermore, deviations in the alignment of magnetic moments found in DFT for the ground state resulted in an increased system energy, further diverging from the experimentally observed stable configuration, as also shown in supplementary information (supplementary figure 2). Our structural model, constructed using a $2\times2$ supercell with a single intercalated Cr atom, revealed an ordered hexagonal lattice of Cr sites, fully consistent with the expected symmetry and geometry of the hexagonal Brillouin zone (see Fig.\ref{fig1}c).\\

The ARPES and DFT calculations, foundational to resolving the system's ground state, are illustrated in Fig.\ref{fig1}. The experimental electronic structure is marked by bands exhibiting pronounced anisotropic orbital character, with significant dispersion along $k_z$, indicative of moderate interlayer coupling and notable overlap of electron wavefunctions along the $c$-axis. By tuning the light polarization between linear horizontal (LH) and linear vertical (LV), we observed distinct changes in the spectral intensity: LH primarily probes out-of-plane orbital components, while LV highlights the in-plane orbital character (Figs.\ref{fig1}d-f). This allows us to visualize the bands of the systems focusing primarily on a certain orbital aspect and, by direct comparison to DFT calculations, gives us extra confidence in benchmarking the features shaping the ground state. For example, around the $A$ point of the Brillouin zone, LH shows the presence of an electron-like band (Fig.\ref{fig1} g), which is also captured by the DFT calculations of Fig.\ref{fig1} i. Such an electron pocket emerges at the $A$ point Fermi level, diminishing progressively with increasing photon energy and disappearing entirely at the $\Gamma$ point. This behavior underscores the system’s electronic structure, blending two-dimensional van der Waals layers with three-dimensional electronic structures.\\

Additionally, variation in photon energy (Fig.\ref{fig1}g and supplementary information figures 4 and 5) reveals minimal shifts in the outer hole bands located at $k$-points near $\pm 0.5$ Å$^{-1}$, consistent with the system’s intrinsic two-dimensional van der Waals nature. In stark contrast, the electronic structure at the center of the Brillouin zone demonstrates a clear three-dimensional character, affirming the system’s "quasi"-van der Waals properties compared to pure CrTe$_{2}$ compounds. This three-dimensionality is especially apparent in the Fermi surface topologies at the $\Gamma$ and $A$ points (Fig.\ref{fig1} h), which differ strikingly. This is further corroborated by the closed contours observed for the fermi surfaces along the $k_z$ direction (supplementary figure 5).\\

\begin{figure}
\centering
    \includegraphics[width=0.9\textwidth]{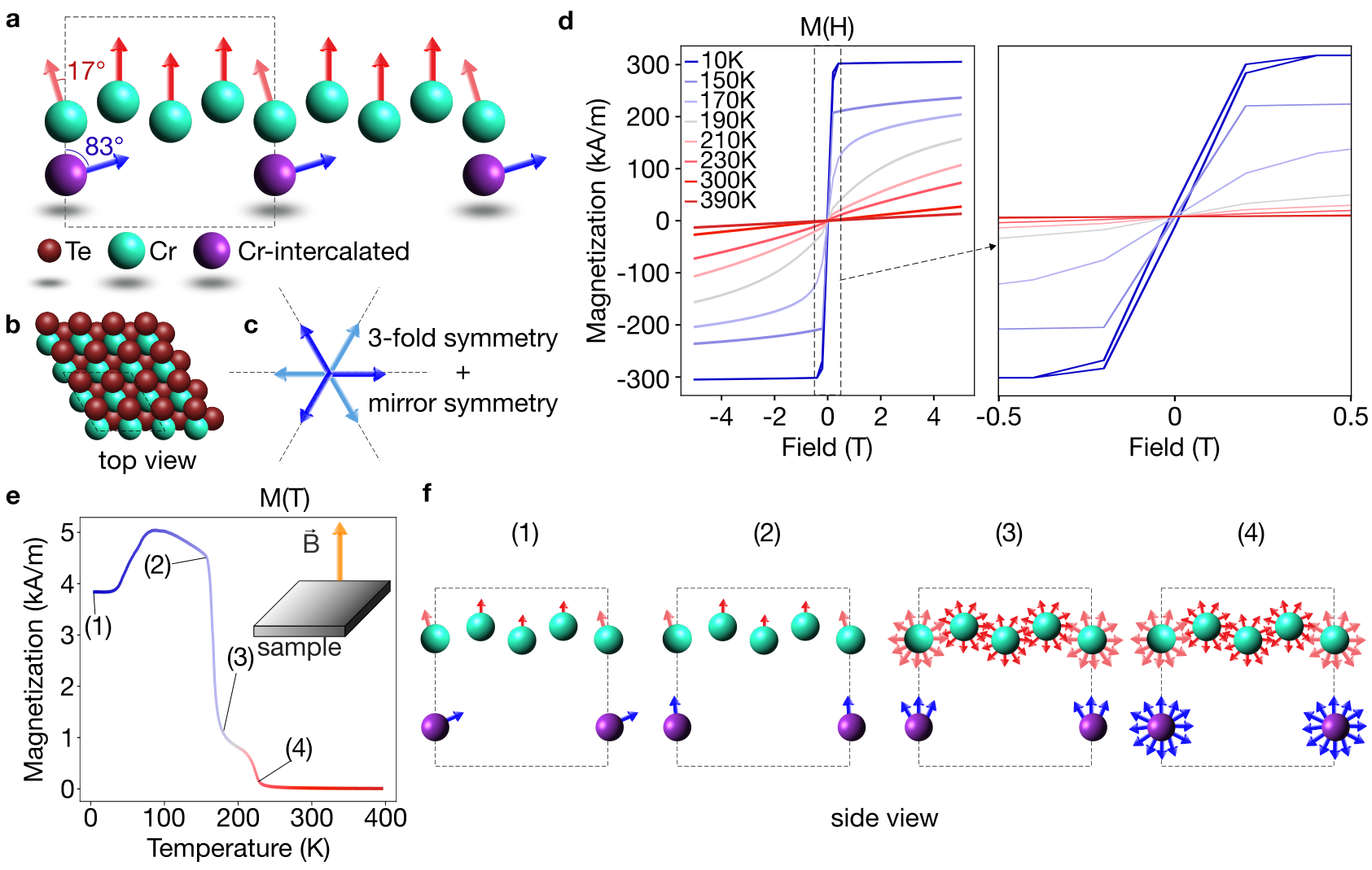}
    \caption{{\bf Magnetic ground state in Cr$_{1.25}$Te$_{2}$} {\bf a.} Side view of the Cr atoms arrangement. Light blue Cr-atoms belong to the CrTe$_2$ layers. Their magnetic moments are aligned ferromagnetically out of plane with a 17$^{\circ}$ canting of the Cr atoms sitting on top of the Cr intercalated within the CrTe$_2$ van der Waals gap (purple color). The magnetic moment of the intercalated atom is nearly transverse to the out of plane direction with a canting of 83$^{\circ}$. 
    {\bf b.} Top view of the crystal structure with the unit cell (dashed lines) and {\bf c.} in-plane spin component of the intercalated Cr resulting in three equivalent minima of the magnetic ground state. {\bf d.} Field dependent magnetization curves measured at different sample temperatures. Magnetic field was applied out of the plane of the sample.  {\bf e.} Magnetization as a function of temperature for an out of plane applied field of 3 mT and {\bf f.} associated sketches of the magnetization mechanism.}
    \label{fig2}
\end{figure}

This behaviour is confirmed by our DFT calculations (Fig.\ref{fig1}i), performed along various high-symmetry directions and at multiple $k_z$ values. The computed band structure aligns closely with experimental data only when the system resides at its minimum energy configuration, leading to the magnetic moment arrangement depicted in Fig.\ref{fig2}a-b. This arrangement features alternating ferromagnetic layers on the atomic scale, with the magnetic moments oriented orthogonally between adjacent planes — a structure fundamentally distinct from full or canted ferromagnetism.\\

The magnetic ground state was verified via spin-ARPES, where spin-resolved measurements across all three spatial components ($S_x$, $S_y$, and $S_z$) were taken at the $\Gamma$ point (Fig.\ref{fig1} j). In adherence to time-reversal symmetry, no net spin polarization should theoretically exist at this point in the absence of magnetic ordering. Nonetheless, our measurements uncover a clear violation of this symmetry, revealing both in-plane and out-of-plane spin components (Fig.\ref{fig1} j), in line with our theoretical predictions for the ground state. These measurements were only possible by using a micro-focused beam (lateral spot size of 10 $\mu$m; see Methods) and likely magnetic domains comparable to the spot size, a hypothesis supported by previous photoemission electron microscopy (PEEM) results \cite{peem1, peem2}. Alternatively, the measurements probed a higher proportion of domains with a consistent orientation. Lastly, we extended our ARPES experiments and DFT simulations to systems with doping levels up to $50\%$ (see Fig.\ref{fig1} k with maps in inverted color scale; see supplementary information for theoretical details, figure 6), arriving at remarkably similar conclusions. These results affirm that orthogonal ferromagnetism is a robust and pervasive feature across a wide doping range.\\

Our findings demonstrate a robust coupling between the intercalated atoms and the CrTe$_2$ layers, as evidenced by the DFT-derived magnetic ground state shown in Fig.~\ref{fig2}a-b. In contrast to other studies, we observe a heterostructure-like orthogonal magnetic arrangement. However, the intercalated Cr atoms exhibit 17$^\circ$ tilt from their planar configuration. This deviation is mirrored in the magnetic moments of adjacent Cr atoms, situated directly above and below the intercalated atoms in neighboring layers (one atom every four), which tilt relative to their vertical orientation by the same degree. This tilting disrupts local symmetry, introducing magnetic frustration within the CrTe$_2$ layers. The energy difference between the out-of-plane spins in the CrTe$_2$ layers and the in-plane spins in the intercalated Cr atoms is calculated to be 8 meV per unit cell. This pronounced magnetic anisotropy is likely driven by the spin-orbit coupling (SOC) of the heavy, non-magnetic Te atoms. The computed total magnetization vector is $[2.8, 0.0, 11.1] \mu_B/\mathrm{fu}$, with individual Cr atoms exhibiting local magnetic moments of approximately 2.8 $\mu_B$. Utilizing the classical spin model, $E=-\frac{1}{2}\sum_{ij}J_{ij}\mathbf{u}_i \cdot \mathbf{u}_j$, where $\mathbf{u}_i$ represents the unit vector of the spin direction for Cr atom $i$ and $J_{ij}$ denotes the exchange interaction energy, we calculate an interaction energy of $J_\mathrm{int} = 4$ meV between neighboring intercalated Cr atoms. Considering six nearest-neighbor interactions, we estimate a mean-field Curie temperature of approximately 90 K for magnetic ordering among the intercalated atoms. However, a more detailed analysis is necessary to provide a rigorous quantitative assessment \cite{Durhuus2023}, which lies beyond the scope of this study. Furthermore, we determine that the nearest-neighbor exchange interaction between the intercalated Cr atoms and those in the CrTe$_2$ layer is $J_1 = -106$ meV, while the next-nearest interaction is $J_2 = 8$ meV. This results in an effective in-plane antiferromagnetic exchange ($J_\mathrm{eff} = J_1 + 6J_2 < 0$) between the intercalated atoms and the CrTe$_2$ layers. The observed preference for transverse magnetic alignment is attributed to long-range interactions between the intercalated atoms and the CrTe$_2$ layers, further enriching the understanding on the magnetic behavior in this system.\\

Orthogonal-ferromagnetic systems with antiferromagnetic exchange present natural heterostructures, exhibiting pronounced anisotropic thermodynamic behavior when comparing in-plane and out-of-plane magnetic alignments. To explore this anisotropy, we examine the temperature and magnetic-field dependencies along these orthogonal directions. For fields applied out-of-plane, SQUID magnetometry reveals that below the magnetic ordering temperature, the magnetization curves (Fig.\ref{fig2}d) display square hysteresis loops, characterized by a small coercive field of 13 mT and a remanence approximately 7$\%$ of the saturation magnetization. This behavior is typical of systems with an easy out-of-plane magnetic axis with weak nucleation field. The effective magnetic anisotropy field is estimated at 0.6 MJ/m$^3$ at 150 K and 0.4 MJ/m$^3$ at 170 K, comparable to values observed in perpendicular magnetic anisotropy materials such as Fe$_3$GeTe$_2$ (0.8 MJ/m$^3$ \cite{tan2018hard}) and CoFeB/MgO (0.5 MJ/m$^3$ \cite{ikeda2010perpendicular}). From the saturation magnetization, the magnetic moment per Cr atom is calculated to be 2.2 ± 0.3 $\mu_B$, lower than the theoretical value of 2.8 $\mu_B$. This discrepancy likely reflects a lower bound, given that full saturation may not be achieved even at 5 Tesla. Additionally, paramagnetic slopes are observed at temperatures exceeding 240 K, indicative of the transition from ferromagnetic to paramagnetic behavior.\\

Further insights into the out-of-plane ferromagnetic order are gained by plotting the temperature-dependent magnetization under a fixed out-of-plane magnetic field of 3 mT, capturing the system's remanent state after low-temperature saturation at 5 Tesla (Fig.\ref{fig2}e). As shown in Fig.\ref{fig2}f: (1) at low temperatures (10–20 K), Cr$_{1.25}$Te$_2$ retains its out-of-plane magnetic configuration, consistent with the schematic in Fig.\ref{fig2}a. As the temperature increases, (2) the magnetization exhibits a broad peak, signaling a weakening of the magnetic order. The applied magnetic field forces the moments, including those of intercalated Cr atoms, to align with the field. At higher temperatures, a double-hump structure appears in the magnetization curve, reflecting the progressive loss of ferromagnetic order and the emergence of paramagnetic behavior: (3) initially, the in-plane Cr moments disorder, leading to a sharp drop in magnetization. (4) Subsequently, the intercalated Cr moments randomize, resulting in paramagnetism above 240 K.\\

\begin{figure}
\centering
    \includegraphics[width=\textwidth]{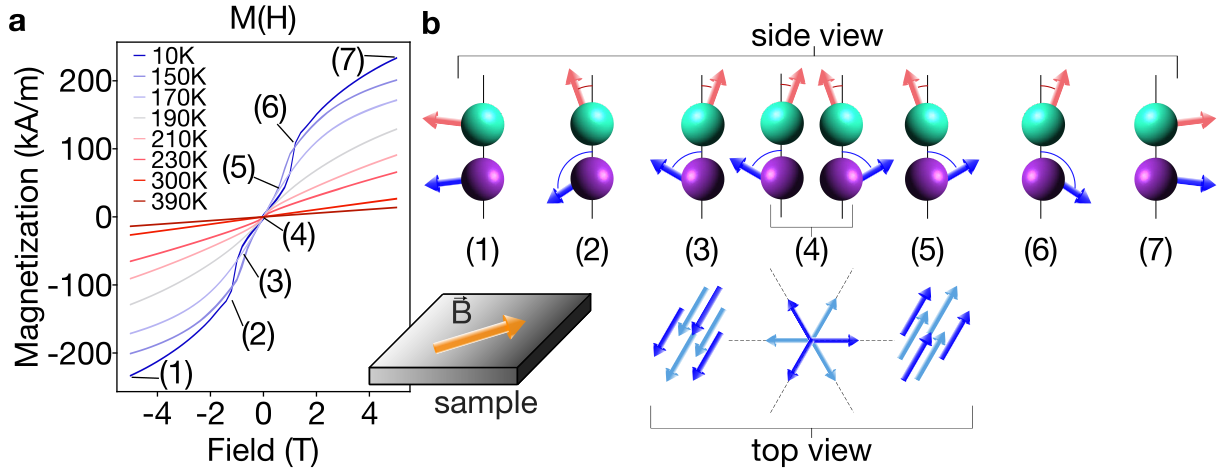}
    \caption{{\bf in-plane magnetization} {\bf a.} Field dependent magnetization curves of Cr$_{1.25}$Te$_{2}$ measured at different sample temperatures and the magnetic field applied along the plane of the sample. The non-saturating slopes of the magnetisation versus magnetic field curves is a typical sign of antiferromagnetic character. Below the magnetic ordering temperature several intermediate steps of the spin reorientation process are described in {\bf b.} Sketch of the spin reorientation mechanism.}
    \label{fig3}
\end{figure}

Along the in-plane direction, we observe a more complex and unprecedented magnetic behavior in these systems. Notably, the magnetization does not reach saturation, even at low temperatures (Fig.\ref{fig3}a), and the curves (spanning 10 K to 150 K) exhibit a distinct change in convexity, manifesting as a characteristic ‘kink’. This feature signals a sudden spin state reorientation, where the magnetic moment reverses direction, shifting from an upward to a downward configuration. This behavior sharply contrasts with previous reports on this compound, where the spins were believed to undergo a smooth reorientation with no abrupt transitions as a function of the applied magnetic field \cite{PhysRevB.100.245114}.While earlier studies struggled to explain a downturn in magnetization and deemed it inconsistent with a gradual reorientation, the abrupt changes observed in our data now reconcile this feature, revealing the true nature of the spin dynamics (see also additional data in supplementary figures 8 and 9).\\

To clarify the observed spin reorientation, we refer to the schematic in Fig.\ref{fig3}b. For simplicity, we begin by considering the ground state configuration, labeled '4', under a positive in-plane magnetic field. A similar process occurs for a negative magnetic field, following a trajectory from '3' to '1'. In the ground state (4), the Cr magnetic moments exhibit a tilt, as previously described. The in-plane magnetic components, primarily contributed by the intercalated Cr atoms, show hexagonal symmetry due to the interplay between rotated magnetic domains and the presence of three energy minima (indicated by blue arrows in the top view). Upon applying a magnetic field (5), the in-plane magnetic moments align with the field direction, resulting in the formation of a single magnetic domain. This alignment corresponds to the first spin reorientation and is reflected in the initial slope change of the magnetization curves in Fig.\ref{fig3}a. As the magnetic field increases, (6) all magnetic domains fully align with the applied field, and a spin-flop-like transition occurs, producing a second slope in the magnetization versus magnetic field curves. Finally, (7) all spins rotate continuously to align with the field direction. This double spin reorientation, reminiscent of spin-flop dynamics, is also observed in Cr$_{1.50}$Te$_2$, extending these magnetic properties to systems with up to 50$\%$ Cr doping (supplementary figure 10). Moreover, this behavior stands in contrast to earlier observations, where Cr vacancies were believed to play a critical role in shaping the magnetic response. Here, we propose that the high crystallinity and large domain sizes—further corroborated by the sharp, well-defined ARPES data—reposition this system as a promising candidate for spin-flop generation.\\

In summary, this study unveils a novel magnetic ground state in Cr$_{1+\delta}$Te$_2$ (with $\delta = 0.25 - 0.50$) that we term orthogonal ferromagnetism, characterized by alternating in-plane and out-of-plane ferromagnetic atomic layers coupled via antiferromagnetic exchange. Our comprehensive multimodal approach, integrating SQUID magnetometry, spin-ARPES, and DFT calculations, provides definitive evidence for a previously overlooked magnetic ordering and resolves ambiguities surrounding the material's magnetic behavior. We observed abrupt spin-flop transitions, in stark contrast to previous reports of gradual reorientation, and place Cr$_{1+\delta}$Te$_2$ as a compelling candidate for spin-flop generation. The robust nature of orthogonal ferromagnetism across varying doping levels underscores its significance in the exploration of emergent quantum phases in low-dimensional materials.\\

\textbf{Acknowledgements}\\
The authors acknowledge M. Brissaud for assistance during some photoelectron spectroscopy measurements. F.M. greatly acknowledges the SoE action of PNRR, number SOE\_0000068 and the funding by the European Union – NextGenerationEU, M4C2, within the PNRR project NFFA-DI, CUP B53C22004310006, IR0000015. This work was supported by the National Research Foundation of Korea (NRF) funded by the Ministry of Education, Science and Technology (NRF-2019M2C8A1057099 and NRF-2022R1I1A1A01063507). J. A. M. acknowledges support from DanScatt (7129-00011B). H. T and T. J acknowledge the support of the French National Research Agency (ANR) (MOSAICS project, ANR-22-CE30-0008). This work was also supported by ANR through the France 2030 PEPR SPIN government grant ANR-22-EXSP 0007.

\textbf{Methods}\\

\textbf{Growth:} To obtain high-quality single crystals, the starting elements were put into a quartz ampule with a capillary bottom. The ampule was evacuated and sealed under a pressure of $2\times 10^{-7}$Torr and placed in a furnace. The furnace was raised to 600 $^\circ$C and held for 72 hours. The furnace temperature was then raised to 1150$^\circ$C at a rate of 10$^\circ$C/h and soaked for 48 hours. For promote the single crystalline growth, the temperature was slowly cooled to room temperature over 10 days.\\

\textbf{Photoelectron Spectroscopy:} Photoelectron spectroscopy experiments were performed at the CX2 endstation of CASSIOP\'EE beamline, SOLEIL synchrotron. The samples were cleaved in ultrahigh vacuum (pressure $1\times10^{-10}$ mbar) and at the temperature of $20$ K, which was kept constant throughout the data acquisition. Linearly ploarized light (both horizontal and vertical) was varied between $20-90$ eV photon energies. Angularly-resolved photoemission spectra were collected by a Scienta R4000 analyzer. The energy and momentum resolutions were better than $15$ meV and $0.08$ \AA$^{-1}$, respectively. The 6-axis manipulator allowed for fine alignment of the sample's high symmetry directions along the analyzer's slit.\\

\textbf{SQUID measurements:} SQUID measurements were performed using a Quantum Design magnetic property measurement system MPMS®3 following standard procedures. Magnetic field sweeps were made in no-overshoot, persistent mode. The curves shown here have been corrected from their diamagnetic signal at high fields.\\

\textbf{DFT calculations:} The DFT calculations were carried out with the electronic structure package GPAW \cite{Enkovaara_2010,Mortensen2024} using plane waves and the projector-augmented wave method. For all calculations we used a plane wave cutoff of 800 eV, a uniform $k$-point grid of $11\times11\times13$ and a Fermi-Dirac smearing of 10 meV. The structure ($2\times2\times1$ repetition of the CrTe$_2$ unit cell with a single intercalated atom) was relaxed until all forces were below 0.005 eV/{\AA} using the PBE functional in a collinear magnetic configuration with intercalated atoms being antialigned with ferromagnetic CrTe$_2$ planes. All subsequent calculations were carried out using the LDA functional and non-collinear DFT including self-consistent spin-orbit. We tested a wide range of different initial non-collinear configurations and found the one shown in Fig. \ref{fig2} to be the most stable. 

All calculated band structures were obtained using the PBE functional and unfolded to the primitive unit cell of CrTe$_2$ \cite{Popescu2012}. Spin-orbit coupling was included non-self consistently in these calculations \cite{Olsen_2016}.\\

\textbf{Raman spectroscopy:} For the Raman spectroscopy measurement the wavelength used was 632.81 nm with 10\% of the laser power. the objective used was x100 with a grating of 600.

\bibliography{biblio}{}

\clearpage

\renewcommand{\thefigure}{S\arabic{figure}}
\setcounter{figure}{0}
\textbf{Supplementary Information: Bilayer orthogonal ferromagnetism in CrTe$_2$-based van der Waals system}

\begin{figure}[hb!]
\centering
    \includegraphics[width=0.9\textwidth]{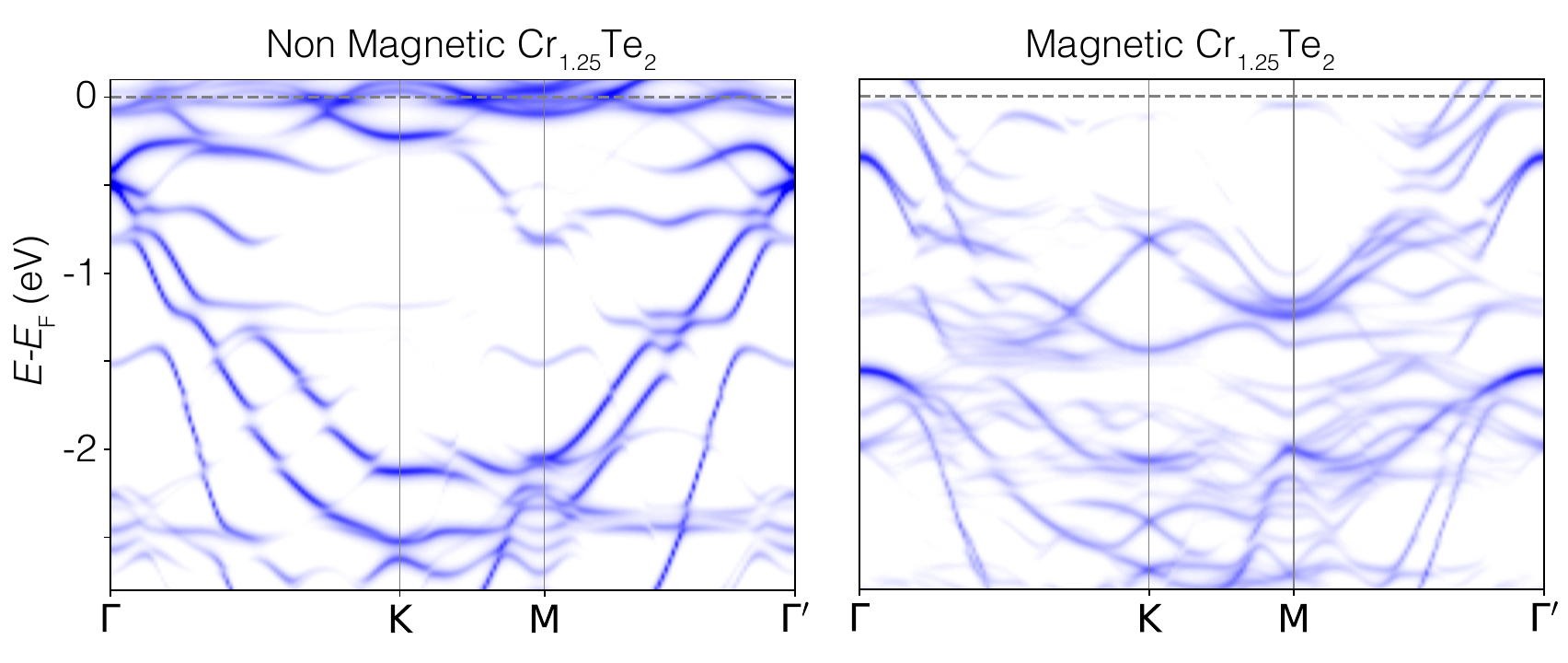}
    \caption{\textbf{Cr$_{1.25}$Te$_2$ with and without magentism.} (Left) non magnetic structure of Cr$_{1.25}$Te$_2$ compared to the (Right) magnetic counterpart. Clear differences are visible not only in terms of the general shape of the electronic structure, but a completely different spectral weight distribution is observed.}
    \label{figS11}
\end{figure}

To underscore the role of magnetism in the electronic structure, we performed DFT calculations for Cr$_{1.25}$Te$_2$ in its non-magnetic ground state. As illustrated in Fig.\ref{figS11}, the non-magnetic band dispersion reveals the presence of prominent, less dispersive bands near the Fermi level, likely shifted above the Fermi level upon the inclusion of magnetic order. These bands, however, are absent in the magnetic state. Notably, our experimental data does not exhibit such dispersions and instead demonstrates a much stronger alignment with the magnetic DFT calculations, emphasizing the significance of magnetic interactions in shaping the system's electronic properties.\\

\begin{figure}
\centering
    \includegraphics[width=0.8\textwidth]{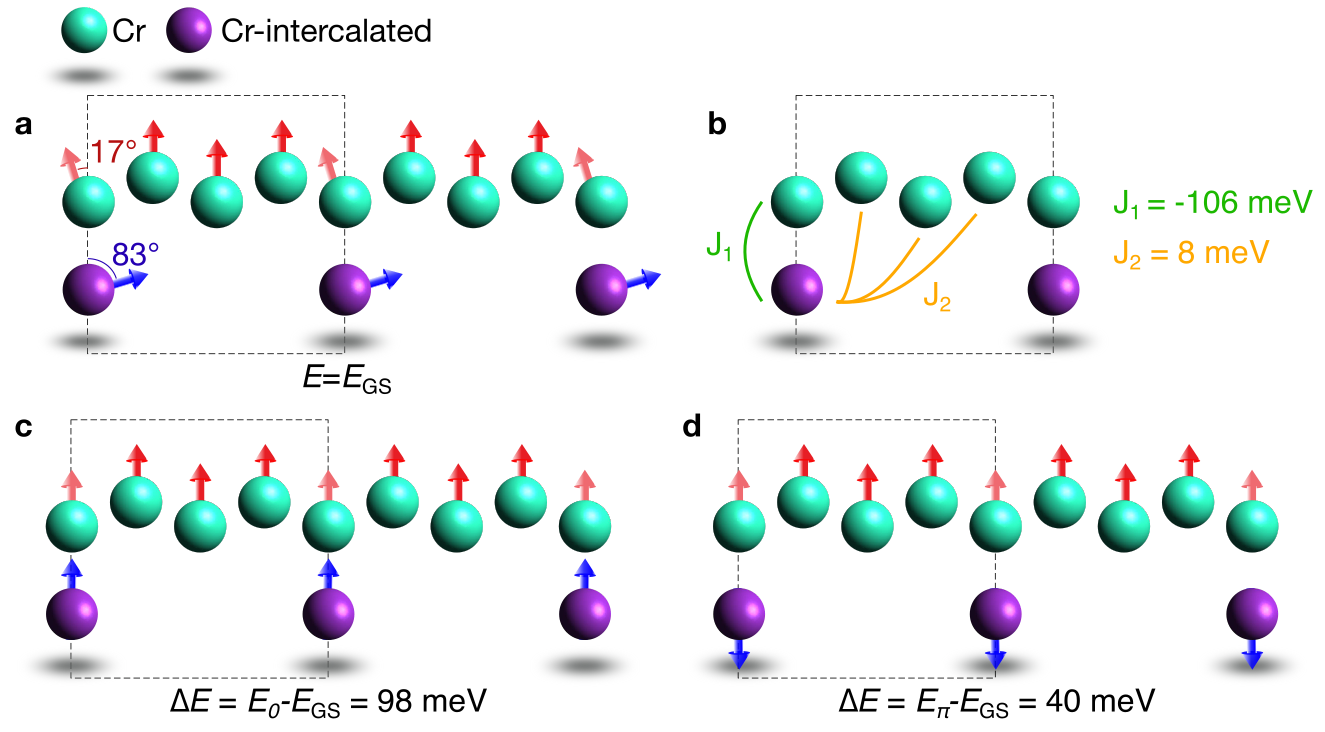}
    \caption{Magnetic configurations tested with first principle calculations \textbf{a.} non-collinear configuration, corresponding to the magnetic groundstate of Cr$_{1.25}$Te$_2$ \textbf{b.} scheme of the nearest-neighbours and the next-nearest-neighbours exchange interactions J$_i$ considered in this work. \textbf{c.} Ferromagnetic configuration ($\varphi_i=0$) and \textbf{d.} antiferromagnetic configuration ($\varphi_i=\pi$).}
    \label{figS3}
\end{figure}

\begin{figure}
\centering
    \includegraphics[width=0.9\textwidth]{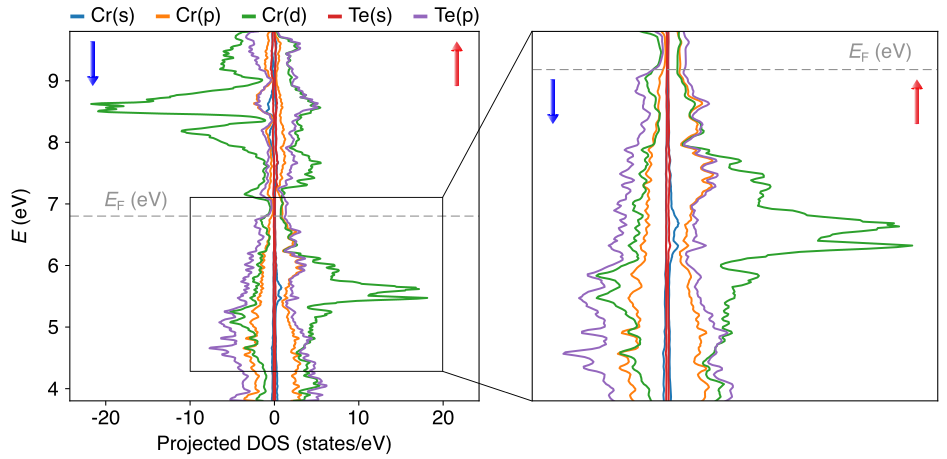}
    \caption{Spin resolved orbital-projected partial DOS.}
    \label{figS4}
\end{figure}

We tested by \emph{ab-initio} calculations a wide range of different initial non-collinear configurations to identify the Cr$_{1.25}$Te$_2$ magnetic groundstate. The calculations were initialized with out-of-plane ferromagnetic order for the Cr in the CrTe$_2$ layers, while the magnetic moment of the intercalated Cr atom was tilted by a starting angle $\varphi_i$. We investigated $\varphi_i=0, \pi/5, 2\pi/5, \pi/2, 3\pi/5, 4\pi/5, \pi$, where $\varphi_i=0$ corresponds to the ferromagnetic alignment configuration. Then the magnetic moments were let free to evolve up to the minimization of the total energy. All calculations ended in the state shown in fig.\ref{figS3}.a and the calculated exchange energies were mapped to a Heisenberg model. The values obtained are antiferromagnetic for the nearest neighbour interlayer Cr-Cr interaction (J$_1=-106$meV), while the next nearest neighbours experienced ferromagnetic coupling (J$_2=8$meV) fig.\ref{figS3}.b. Only two exceptions were obtained for the $\varphi_i=0$ and $\varphi_i=\pi$ starting configurations, which maintained (fig.\ref{figS3}.c) antiferromagnetic and (fig.\ref{figS3}.d) antiferromagnetic alignment, respectively. However, their minimal energy is significantly higher when compared to the non-collinear configuration (i.e. $98$meV and $40$meV for the ferromagnetic and antiferromagnetic alignment, respectively). Therefore, the configuration in fig.\ref{figS3}.a is the most stable and we identified it as the magnetic groundstate of Cr$_{1.25}$Te$_2$. \ref{figS4} reports the spin-resolved projected density of state (p-DOS) showing that Te-p, Cr-d and Cr-p orbitals are the primary contributors to the electronic structure near the Fermi level. \\

\begin{figure}
\centering
    \includegraphics[width=0.9\textwidth]{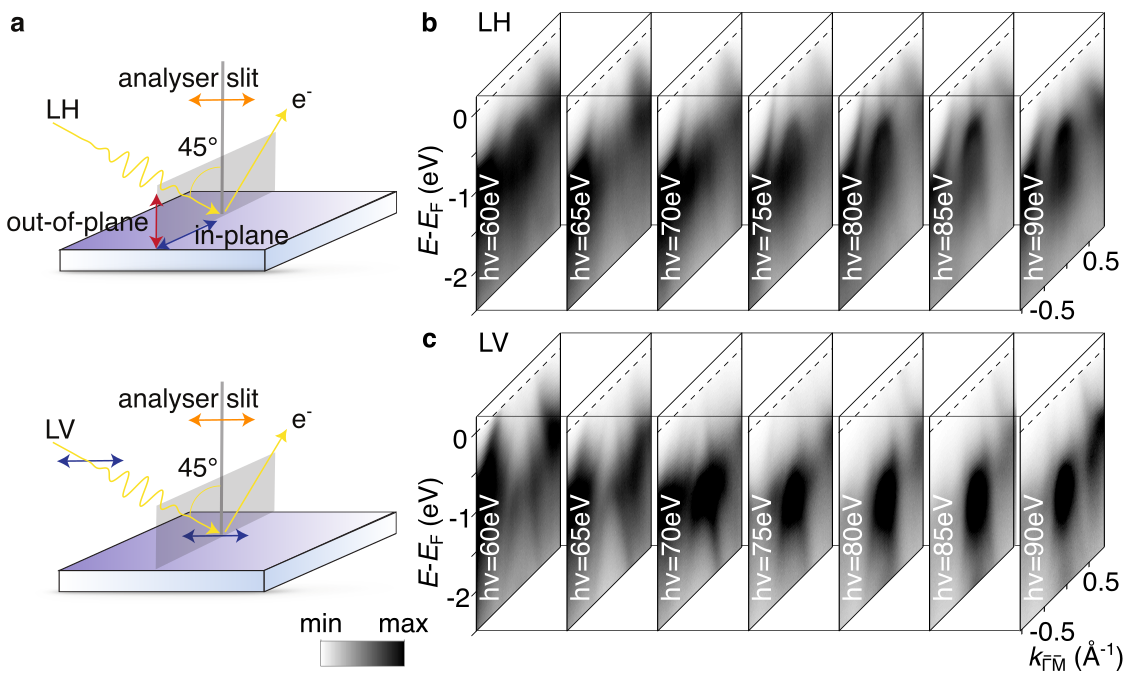}
    \caption{{\bf a.} Experimental geometry and photon energy scans acquired along the $\overline{\Gamma}-\overline{M}$ direction for both {\bf b.} linear horizontal and {\bf c.} linear vertical light polarization.}
    \label{figS5}
\end{figure}

\begin{figure}
\centering
    \includegraphics[width=0.7\textwidth]{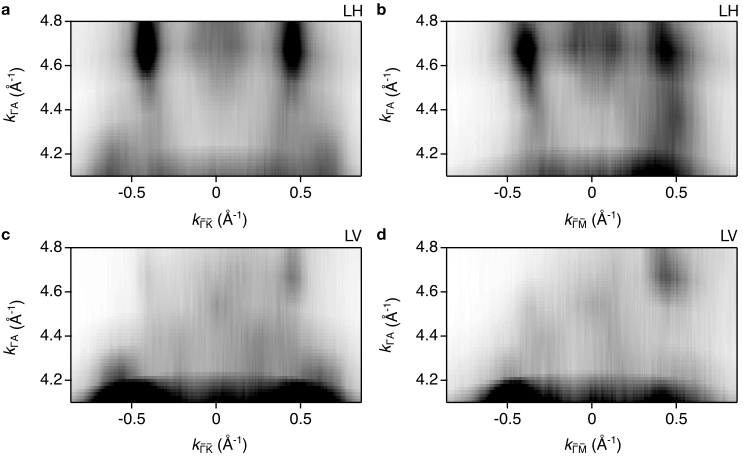}
    \caption{$k_z$ \emph{versus} $k_{||}$ obtained by scanning h$\nu$ from 60 eV to 90 eV with both linear horizontal (\textbf{a., b.}) and vertical (\textbf{c., d.}) polarization. Sample is oriented with the {\textbf{a., c.}} $\overline{\Gamma}-\overline{K}$ direction and {\textbf{b., d.}} $\overline{\Gamma}-\overline{M}$ direction along the analyser slit.}
    \label{figS6}
\end{figure}

We performed photoelectron spectroscopy experiments by using linearly polarized light, both horizontal (LH) and vertical (LV). The analyser slit is oriented vertically in our experimental geometry (\ref{figS5}.a) in such a way that the LV polarization will be within the surface plane of the sample and parallel to the analyser slit, uniquely probing in-plane orbitals. On the other hand LH polarization will couple to both in-plane and out-of-plane orbitals with equal contribution ($50$\% each) thanks to the $45^\circ$ impinging angle of the photon beam. As seen in \ref{figS5}.b and \ref{figS5}.c, the photemitted signal is strongly dichroic and different electronic states are highligten/suppressed when probed with LH/LV polarized light. Nevertheless, both polarizations give strong photemission intensity, thus indicating a mixed in-plane/out-of-plane orbital character at the Fermi level. The electronic dimensionality at the Fermi level is determined by the $k_z$ \emph{versus} $k_{||}$ plots reported in \ref{figS6}. The outer states at $k_{||}\sim\pm0.5$ \AA are rather stripy, signature of their mostly two-dimensional character similarly to other purely van der Waals systems. Conversely, Cr$_{1.25}$Te$_2$ $k_z$ plots do also show an higly-dispersive, closed contour centred at the A point, while it is completely absent at the Brillouine zone centre ($\Gamma$). Such a strong three-dimensional state is a direct consequence of the intercalated Cr atoms which increase the overlapping of the out-of plane orbitals in Cr$_{1.25}$Te$_2$. This is also hinted by the strong dichroic response of this state, which is mostly visible when probed by LH polarization (\ref{figS5} and \ref{figS6}).

\begin{figure}
\centering
    \includegraphics[width=0.9\textwidth]{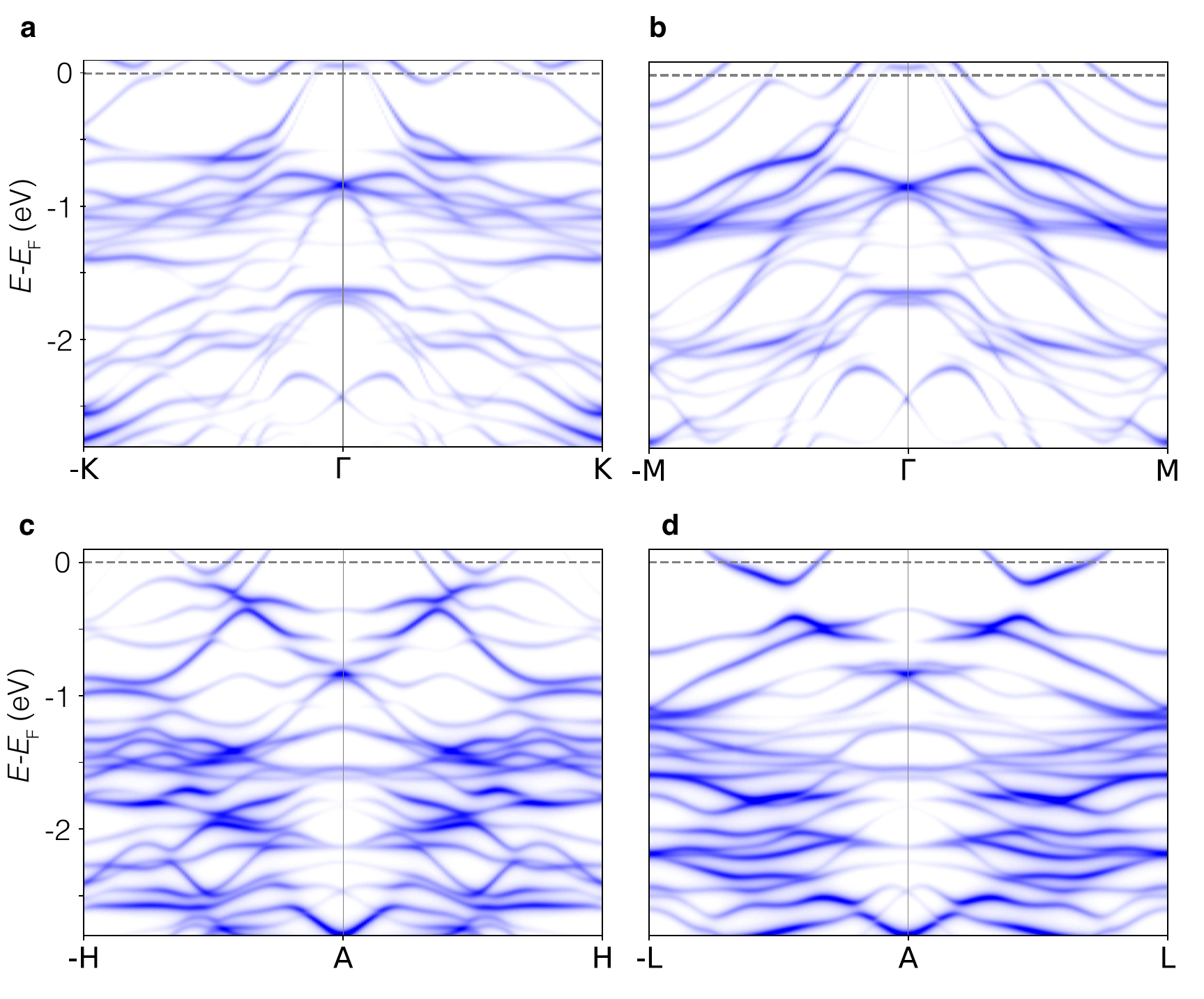}
    \caption{\textbf{Cr$_{1.5}$Te$_2$ DFT calculations.} Calculations for $50\%$ of Cr intercalation shown along the high symmetry directions, \textbf{a.} $\Gamma-K$, \textbf{b.} $\Gamma-M$, \textbf{c.} $A-H$ and \textbf{d.} $A-L$}
    \label{figS12}
\end{figure}

We have also performed DFT calculations along the high symmetry directions for Cr$_{1.5}$Te$_2$ and found a behaviour similar to our experimental results shown in the main text (See Fig.\ref{figS12}).
 
\begin{figure}
\centering
    \includegraphics[width=0.6\textwidth]{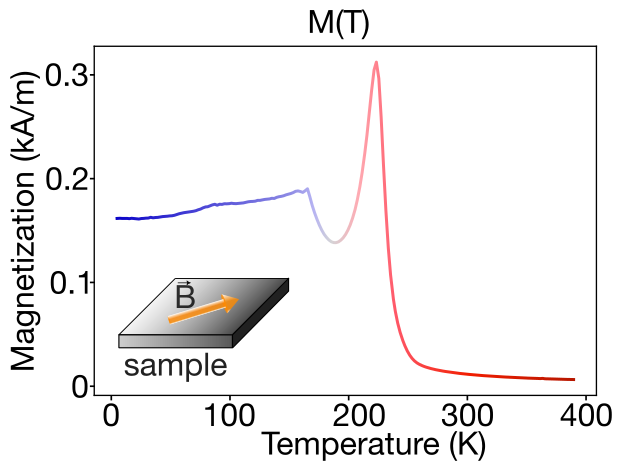}
    \caption{Magnetization as a function of temperature for an in-plane applied field of 3 mT.}
    \label{figS1}
\end{figure}

The temperature magnetization curve in \ref{figS1} shows an in-plane ordering settling slightly below $\sim 230$ K with the appearing of a large magnetization peak. One can observe a lower magnetization value IP rather than OOP, thus a typical antiferromagnetic behaviour with the cusp at the ordering temperature going back down lowering the temperature of acquisition.

\begin{figure}
\centering
    \includegraphics[width=\textwidth]{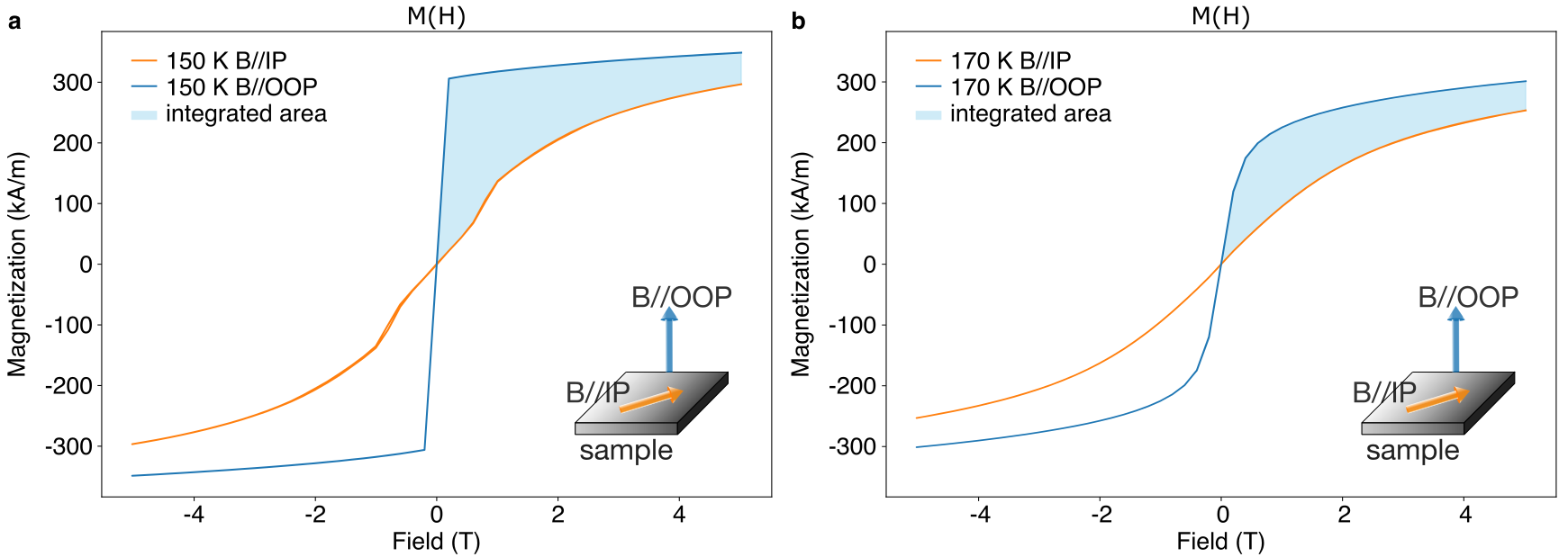}
    \caption{Single-ride magnetization curves IP and OOP at different temperatures used to calculate the perpendicular magnetic anisotropy energy. The extracted values are \textbf{a.} 0.68 MJm$^{-3}$ for the magnetization curves obtained at 150 K and \textbf{b.} 0.45 MJm$^{-3}$ at 170 K, i.e. below and above the spin-reorientation respectively. }
    \label{figS2}
\end{figure}

The perpendicular magnetic anisotropy energy was extracted from single-ride magnetization curves IP and OOP (\ref{figS2}). The area enclosed between the OOP and IP curves, highlighted in blue in \ref{figS2}, allowed us to find the effective perpendicular magnetic anisotropy of the system. Pointing up the OOP Ms value at 5T, one can calculate the uniaxial anisotropy of the system. For this, we removed the calculated shape anisotropy energy ($\mu_0$/2.Ms$^{2}$). We found 0.68 and 0.45 MJm$^{-3}$ at 150, 170K respectively (i.e. i.e. below and above the spin-reorientation respectively), confirming once again a large OOP magnetic anisotropy.

\begin{figure}
\centering
    \includegraphics[width=0.9\textwidth]{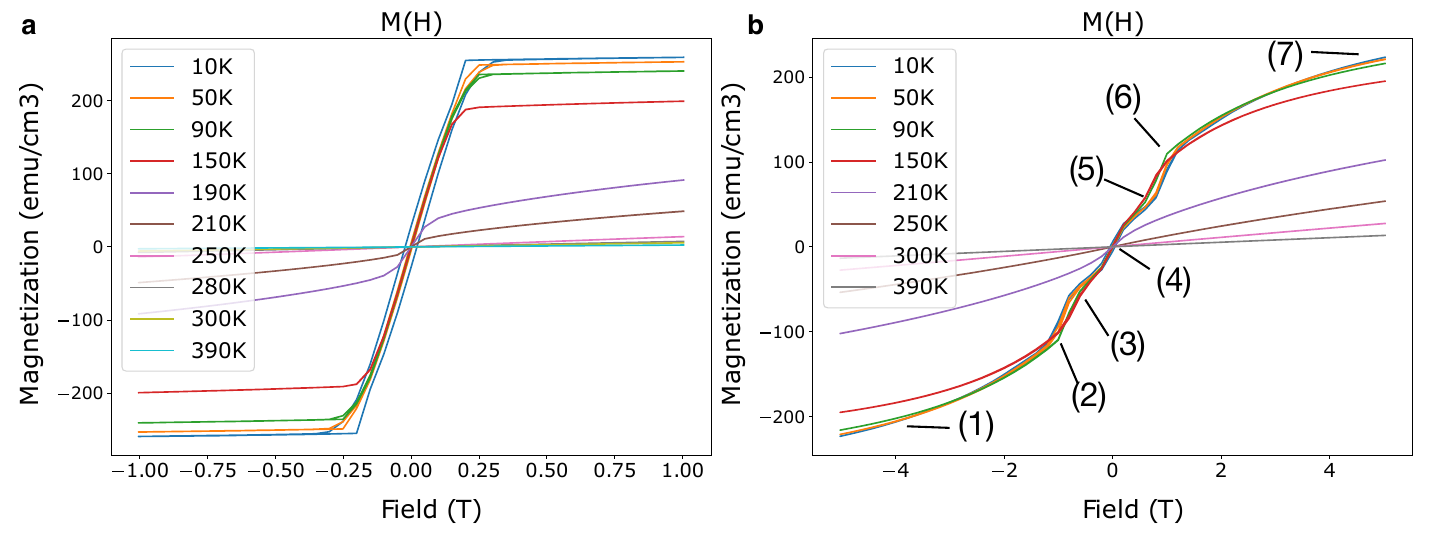}
    \caption{\textbf{Cr$_{1.5}$Te$_2$ sister compound.} \textbf{a.} Field dependent magnetization curves for OOP applied magnetic field for different temperatures. \textbf{b.} Field dependent magnetization curves for magnetic field in IP configuration. \textbf{c.} Fermi surface maps acquired with 86eV and \textbf{d.} 63eV linearly polarized photon energies \textbf{e.} Band dispersion along $\Gamma$-K and both (\emph{left}) linear horizontal and (\emph{right}) linear vertical 86eV photon energy. \textbf{f.} $k_z$ versus $k_{//}$ measured with $\Gamma$-K direction aligned with the analyser slit.}
    \label{figS10}
\end{figure}

To enhance the robustness of our study, we conducted SQUID magnetometry and photoelectron spectroscopy on a related sister compound, Cr$_{1.5}$Te$_2$, featuring 50$\%$ intercalated Cr. Although we observed a slightly different hysteresis loop, our findings unequivocally demonstrate that the underlying physics associated with field-induced spin reorientation remains consistent, as illustrated in Figure \ref{figS10}. This crucial observation underscores the persistence of this phenomenology across a broad spectrum of Cr intercalants, further validating the generality of our results.

\end{document}